\documentclass[a4paper,11pt]{article}
\usepackage{pos}
\usepackage{natbib}
\usepackage[utf8]{inputenc}
\usepackage{amsmath}

\title{Radial Oscillations in Hybrid Stars with Slow Quark Phase Transition}

\author*[a,b]{Ishfaq Ahmad Rather}
\author[c]{Kauan D. Marquez}
\author[d]{Betânia C. Backes}
\author[e]{Grigoris Panotopoulos}
\author[b]{Il{\'i}dio Lopes}

\affiliation[a]{Institut f\"{u}r Theoretische Physik, Goethe Universit\"{a}t, 60438 Frankfurt am Main, Germany}
\affiliation[b]{CENTRA, Instituto Superior T{\'e}cnico,
Universidade de Lisboa, 1049-001 Lisboa, Portugal}

\affiliation[c]{CFisUC, Departmento de Física, Universidade de Coimbra, 3004-516 Coimbra, Portugal}
\affiliation[d]{School of Physics, Engineering and Technology, University of York, YO10-5DD York, United Kingdom}
\affiliation[e]{Departamento de Ciencias F{\'i}sicas, Universidad de la Frontera, Casilla 54-D, 4811186 Temuco, Chile}

\emailAdd{rather@astro.uni-frankfurt.de}
\emailAdd{marquezkauan@gmail.com}
\emailAdd{betania.backes@york.ac.uk}
\emailAdd{grigorios.panotopoulos@ufrontera.cl}
\emailAdd{ilidio.lopes@tecnico.ulisboa.pt}

\abstract{This study investigates the radial oscillations of hybrid neutron stars, characterized by a composition of hadronic external layers and a quark matter core. Utilizing a density-dependent relativistic mean-field model that incorporates hyperons and
baryons for describing hadronic matter, and a density-dependent quark model for quark matter, we analyze the ten lowest eigenfrequencies and their corresponding oscillation functions. Our focus lies on neutron stars with equations-of-state involving N, N + $\Delta$, N + H, and N + H + $\Delta$, featuring a phase transition to quark matter. Emphasizing the effects of a slow phase transition at the hadron-quark interface, we observe that the maximum mass is attained before the fundamental mode's frequency decreases for slow phase transitions. This observation implies the stability of stellar configurations with higher central densities than the maximum mass, called Slow Stable Hybrid Stars (SSHSs), even under small radial perturbations. The length of these SSHS branch depends upon the energy density jump between two phases and the stiffness of the quark EoS.}

\FullConference{%
  10th International Conference on Quarks and Nuclear Physics (QNP2024)\\
  8-12 July, 2024\\
Barcelona, Spain
}

\begin{document}
\maketitle

\section{Introduction}
\label{intro}
Neutron stars, with radii around 10 km and masses up to twice that of the Sun, contain the densest matter in the universe, making them ideal for studying cold, dense nuclear matter. Their properties, such as mass and radius, are governed by the nuclear equation of state (EoS), though the exact EoS at densities above nuclear saturation remains uncertain, potentially involving exotic degrees of freedom like hyperons and delta baryons \cite{Glendenning:1997wn, PhysRevC.106.055801}.  Observations, such as gravitational waves (GWs) from events like GW170817 \cite{PhysRevLett.121.161101}, offer critical constraints on the EoS, shedding light on the internal structure of these stars. Radial and non-radial oscillations, excited by various astrophysical processes, play a key role in understanding stability and composition, enabling investigations into the star's internal constituents and a wide array of its properties \cite{1964ApJ...140..417C, 1967ApJ...149..591T, PhysRevD.107.123022}. Our study explores radial oscillations in stars transitioning from hadronic to quark matter, incorporating hyperons and delta baryons, using density-dependent relativistic mean-field models. This research highlights the interplay of exotic phases and phase transitions, offering new avenues to probe dense matter properties and detect unique features via multi-messenger astronomy.

\section{Theory and Formalism}
\label{theory}
\subsection{Equation of State}
To study the hadronic matter, we employ the widely used DD-RMF model. This model is very successful and widely used for describing neutron star matter, as it replaces the self- and cross-coupling of mesons in the RMF model with density-dependent nucleon-meson coupling constants. In the present work, we employ DDME2 parametrization \cite{PhysRevC.71.024312} with extension to hyperons and deltas using the approach suggested by \cite{Lopes1}.

The energy density for this parametrization is given by
\begin{align}\label{1a}
\mathcal{E}={}& \sum_b \frac{\gamma_b}{2\pi^2}\int_0^{{k_{F}}_b} dk k^2 \sqrt{k^2 + {m_b^\ast}^2}+ \frac{m_\sigma^2}{2} \sigma_0^2+\frac{m_\omega^2}{2} \omega_0^2 +\frac{m_\phi^2}{2} \phi_0^2  + \frac{m_\rho^2}{2} \rho_{03}^2 ,
\end{align}
where the sum runs over the baryons considered in a given matter composition (nucleons $N=\{n,p\}$, hyperons $H=\{\Lambda,\Sigma^-,\Sigma^0,\Sigma^+,\Xi^-,\Xi^0\}$), and/or particles of the spin-3/2 baryon decuplet such as the deltas $\Delta=\{\Delta^-,\Delta^0,\Delta^+,\Delta^{++}\}$. The pressure is then obtained from the fundamental relation
$
    P =\sum_i \mu_i n_i - \mathcal{E} + n_B \Sigma^r,
$
with a correction from the rearrangement term due to the density-dependent couplings, to guarantee thermodynamic consistency and energy-momentum conservation.

\subsection{Quark Matter}
For the quark matter, we employ the density-dependent quark mass (DDQM) model for describing quark matter due to its simplicity and versatility. This model has been used to explore the deconfinement phase transition \cite{backes2021effects}, making it suitable for our study of hybrid stars.

The energy density can be viewed as the one for a free system with particle masses ${m_i(n_B)}$  and effective chemical potentials ${\mu_i^*}$ and is written as
\begin{equation}
    \mathcal{E} = \Omega_0 (\{\mu_i^*\},\{m_i\}) + \sum_i \mu_i^* n_i,
    \label{free-system-fundamental-eq}
\end{equation}
where ${\Omega_0}$ is the thermodynamic potential of a free system. The expression for the pressure is
\begin{align}
    P ={}& -\mathcal{E} + \sum_i \mu_i n_i
    = -\Omega_0 + \sum_i (\mu_i - \mu_i^*)n_i
    = -\Omega_0 + \sum_{i,j} \frac{\partial \Omega_0}{\partial m_j}n_i\frac{\partial m_j}{\partial n_i},
    \label{pressure-quarks}
\end{align}

To construct the phase transition, we use the Maxwell construction  where
\begin{equation}
\begin{gathered}
  P^{(i)}=P^{(f)}=P_0 ~~ \text{and}~~  \mu^{(i)}(P_0)=\mu^{(f)}(P_0)=\mu_0,
\end{gathered} \label{eq:gibbscon}
\end{equation}
sets the transition between the initial (${i}$) and final (${f}$) homogeneous phases with the total chemical potential written as
$ \mu^{ (i,f)}={(\mathcal{E}^{(i,f)}+P^{(i,f)})}/{n_B^{(i,f)}}$,
where ${\mathcal{E}^{(i,f)}}$, ${P^{(i,f)}}$ and ${n_B^{(i,f)}}$ are the total energy density, pressure, and baryon number density, obtained from the effective models of each phase.

\subsection{Radial oscillations}

The small perturbation of the equations governing the dimensionless quantities $\xi$ = $\Delta r/r$ and $\eta$ = $\Delta P/P$, with $\Delta r$ being the radial displacement and $\Delta P$ being the pressure perturbation, are defined as \cite{1997A&A...325..217G}
{\small
 \begin{align}\label{ksi}
     \xi'(r) ={}& -\frac{1}{r} \Biggl( 3\xi +\frac{\eta}{\gamma}\Biggr) -\frac{P'(r)}{P+\mathcal{E}} \xi(r),\\
\label{eta}
          \eta'(r) ={}& \xi \Biggl[ \omega^{2} r (1+\mathcal{E}/P) e^{\lambda - \nu } -\frac{4P'(r)}{P} -8\pi (P+\mathcal{E}) re^{\lambda} 
     +  \frac{r(P'(r))^{2}}{P(P+\mathcal{E})}\Biggr] 
     + \eta \Biggl[ -\frac{\mathcal{E}P'(r)}{P(P+\mathcal{E})} -4\pi (P+\mathcal{E}) re^{\lambda}\Biggr] ,
 \end{align}
 }
 where $\omega$ is the frequency oscillation mode. $\gamma$ is the adiabatic relativistic index and $c_s^{2}$ is the speed of sound squared.
Supplemented with two additional boundary conditions, at the center, and at the surface, that must be satisfied, the frequencies are then computed by
 \begin{equation}
\nu = \frac{\omega}{2\pi} = \frac{s \: \omega_0}{2\pi} ~~(kHz),
\end{equation}
where $s$ is a dimensionless number, while $\omega_0 \equiv \sqrt{M/R^3}$.
We apply the shooting method, starting with a trial $\omega^2$ and initial conditions that meet the center's boundary requirement. We then integrate towards the surface, and the specific $\omega^2$ values that fulfill the boundary conditions indicate the eigenfrequencies of radial disturbances.

These equations describe the Sturm-Liouville eigenvalue equations for $\omega$. The solutions yield the discrete eigenvalues $\omega_n ^{2}$, arranged in increasing order as \( \omega_0 ^{2} < \omega_1 ^{2} < \, ... \, < \omega_n ^{2} \). The integer $n$ represents the number of nodes for a specific NS. The star is stable if $\omega$ is real, while an imaginary frequency indicates instability. 

However, some studies have examined two extreme scenarios based on conversion timescales: slow and fast \cite{Pereira:2017rmp}. Numerical calculations have shown that for slow conversion, even if $\partial M/{\partial \mathcal{E}_c} < 0$, $\omega_0^2$ can still be a real number, thus implying stability in hybrid stars with sharp density discontinuities and slow interface conversions. They are referred to as slow-stable hybrid stars (SSHSs) \cite{Lugones:2023xeq}. 

\section{Results and Discussion}
\label{results}

\begin{figure}[ht]
\centering
		\begin{minipage}[t]{0.49\textwidth}		 		
  \includegraphics[width=\textwidth]{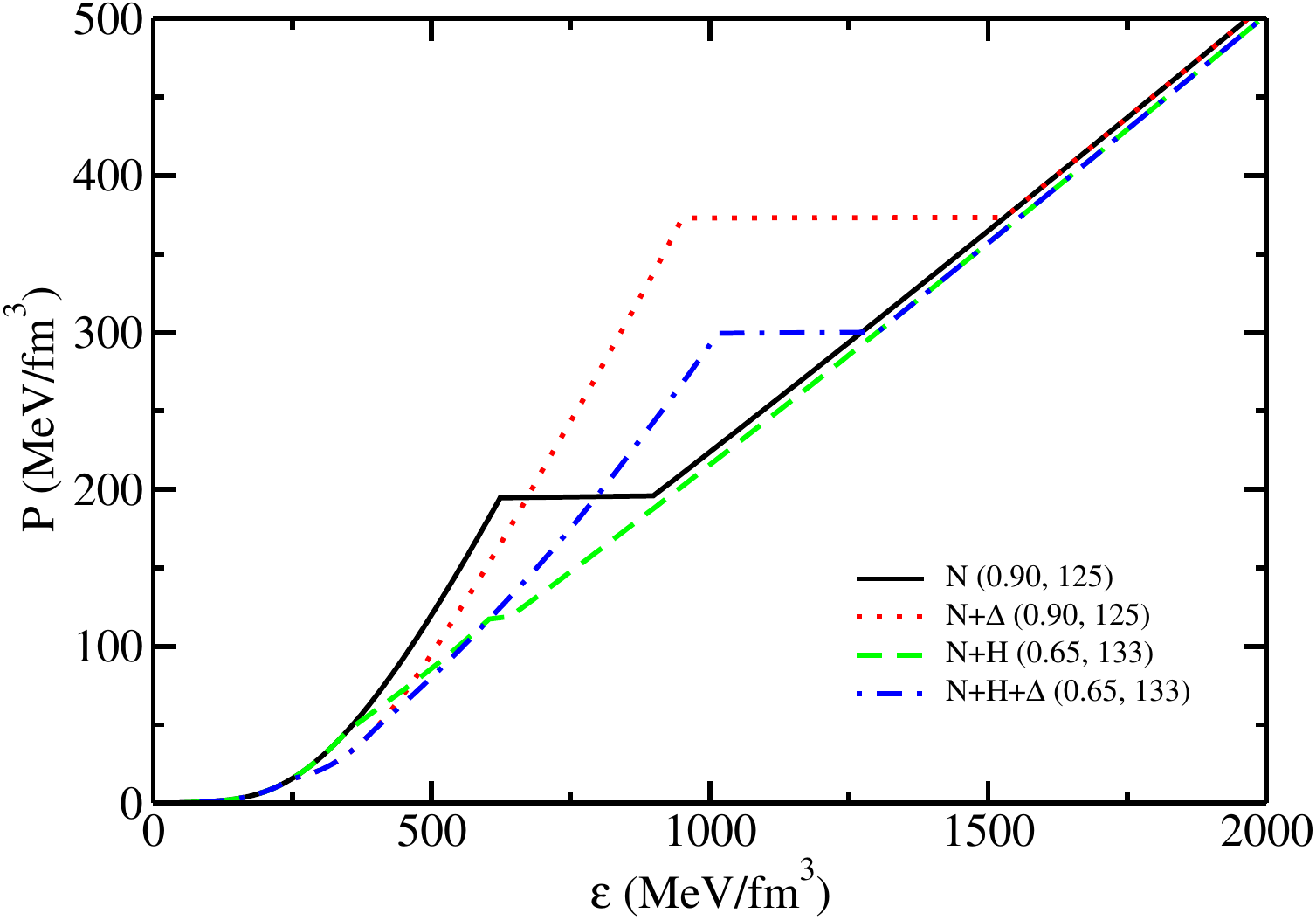}
			 	\end{minipage}
		 		\begin{minipage}[t]{0.49\textwidth}
			 		\includegraphics[width=\textwidth]{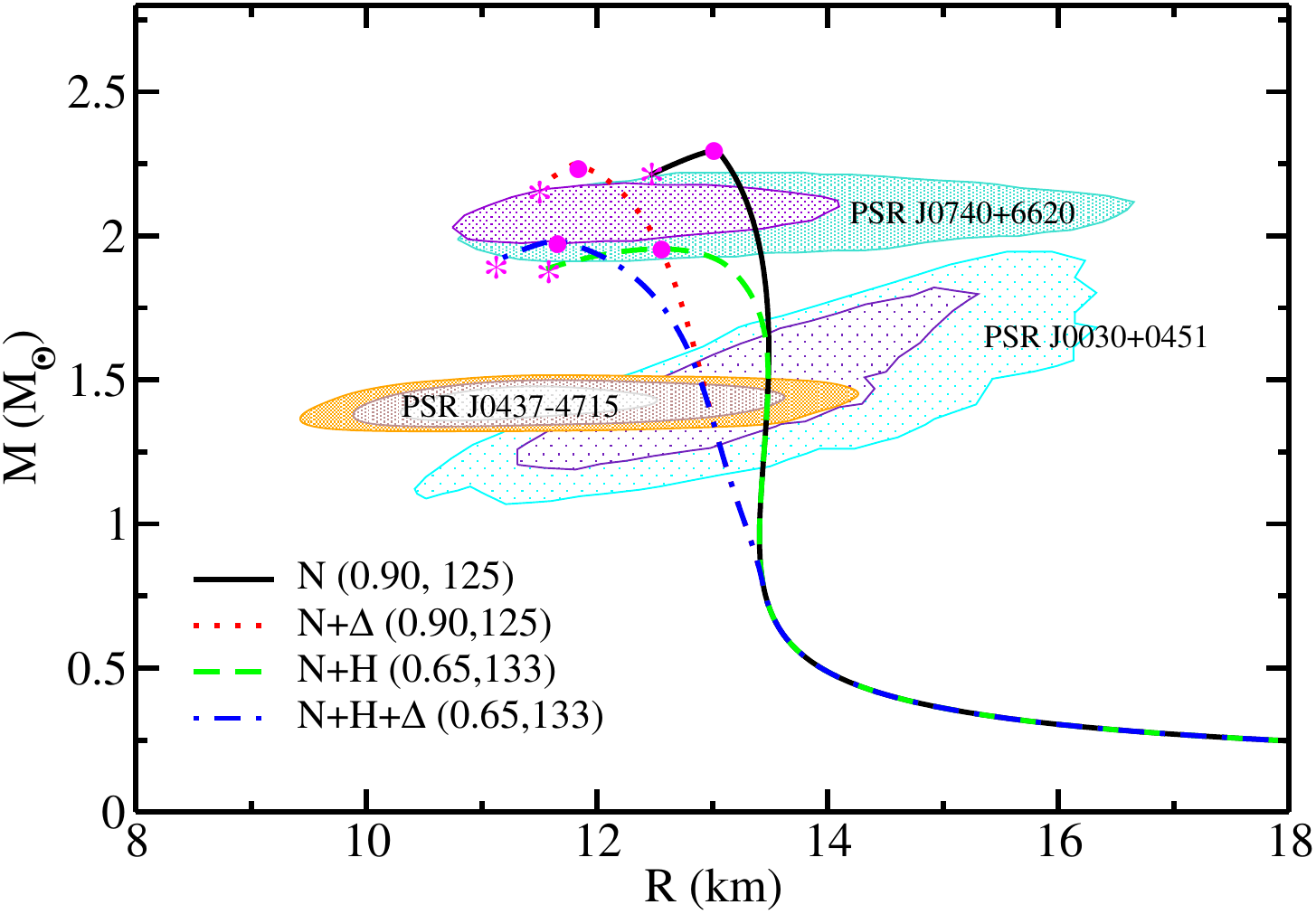}
			 	\end{minipage}
	\caption{Energy density and pressure variation (left) and Mass-Radius relations (right) for different model compositions with a phase transition to the quark matter at different parameter values (${C, D^{1/2}}$). In the right panel, the solid symbol marks the last stable point which is the maximum mass configuration. The line from solid to the star symbol marks the SSHS branch. The several credible regions for mass and radius are inferred from the analysis of PSR J0740+6620, PSR J0030+0451, and PSR J0437-4715 \cite{miller2021,2021ApJ...918L..27R, Miller_2019a, Riley_2019, Choudhury:2024xbk}.}
	\label{fig1} 
\end{figure}

The left panel of Figure \ref{fig1} illustrates the equation of state (EoS), showing pressure versus energy density for neutron stars (NSs) considering various hadronic compositions and parameter sets. Unlike nucleonic-only EoSs, those including exotic particles exhibit unique features: at intermediate densities, the EoS reaches a plateau at the coexistence pressure and stiffens post-phase transition. While pure nucleonic matter initially produces the stiffest EoS, the addition of $\Delta$ baryons softens it due to redistributed Fermi pressure among additional particle species. These $\Delta$ baryons, alongside hyperons, significantly decrease the effective nucleon mass at higher densities, further softening the EoS. The interplay of hyperons and $\Delta$s leads to a soft EoS at lower densities and stiff at higher densities. Beyond the phase transition, the hybrid EoS becomes relatively uniform, with the stiffness primarily influenced by the coexistence point rather than parameter variations, offering insights into hybrid star modeling.

The mass-radius relation for NSs derived from solving the TOV equations is shown in the right panel of Figure \ref{fig1} for various hybrid EoSs. The maximum mass decreases from 2.30 ${M_{\odot}}$ for nucleonic-only (N) to 2.25 ${M_{\odot}}$ when $\Delta$ baryons appear. For N+H and N + H + $\Delta$ EoS, the maximum mass changes from 1.97 to 1.98 ${M_{\odot}}$. All EoSs satisfy the necessary astrophysical constraints. The MR relation also features SSHSs between the maximum mass configuration (marked by dots) and the $f$-mode frequency limit (stars), with the SSHS branch length depending on the energy density jump and quark EoS stiffness. For N+H EoS, the smallest energy density jump (38 MeV/fm$^3$) and stiff quark EoS yield the largest SSHS radius of 0.88 km.

\begin{figure*}[t]
		\begin{minipage}[t]{0.49\textwidth}		 		
  \includegraphics[width=\textwidth]{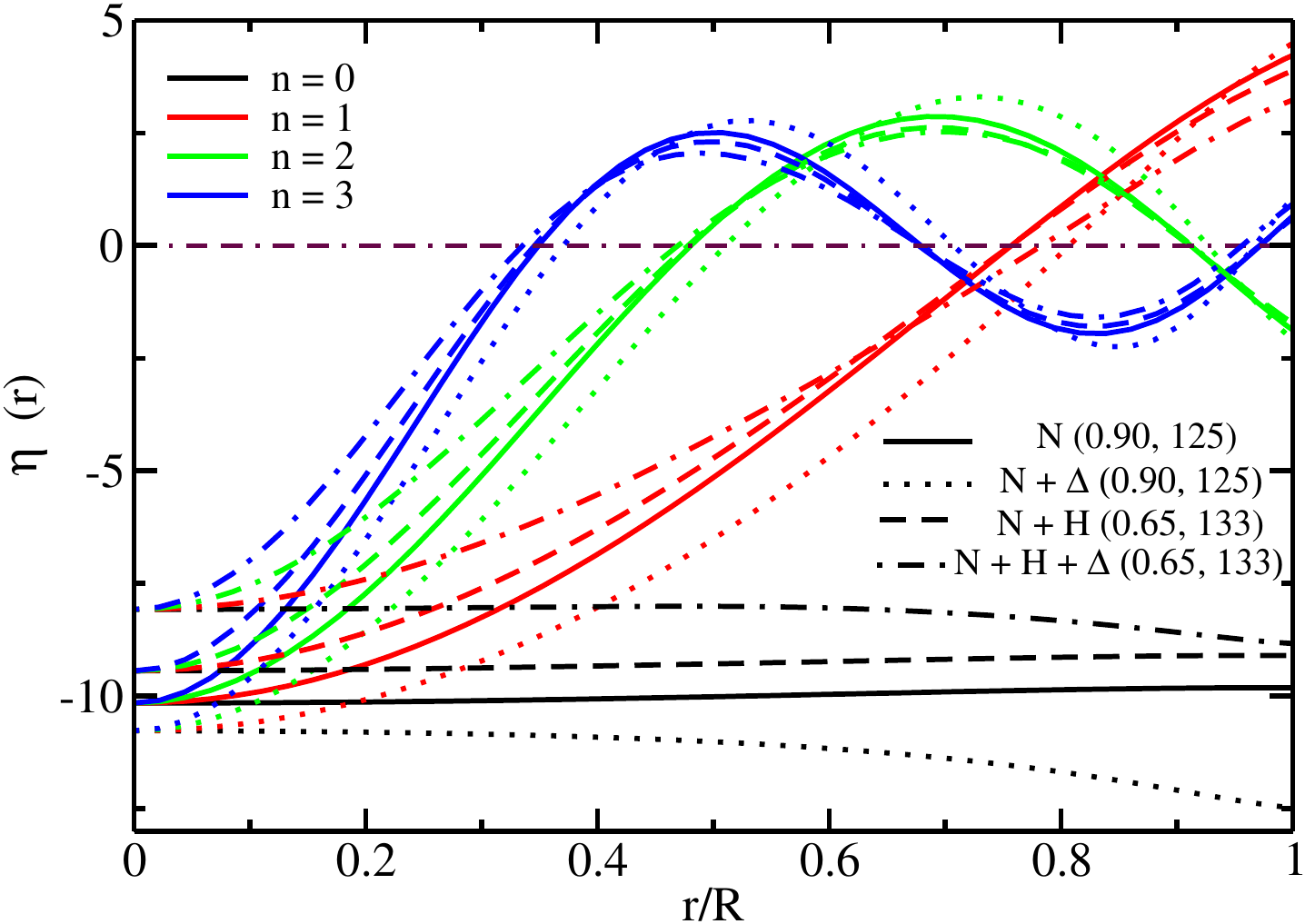}
			 	\end{minipage}
		 		\begin{minipage}[t]{0.49\textwidth}
			 		\includegraphics[width=\textwidth]{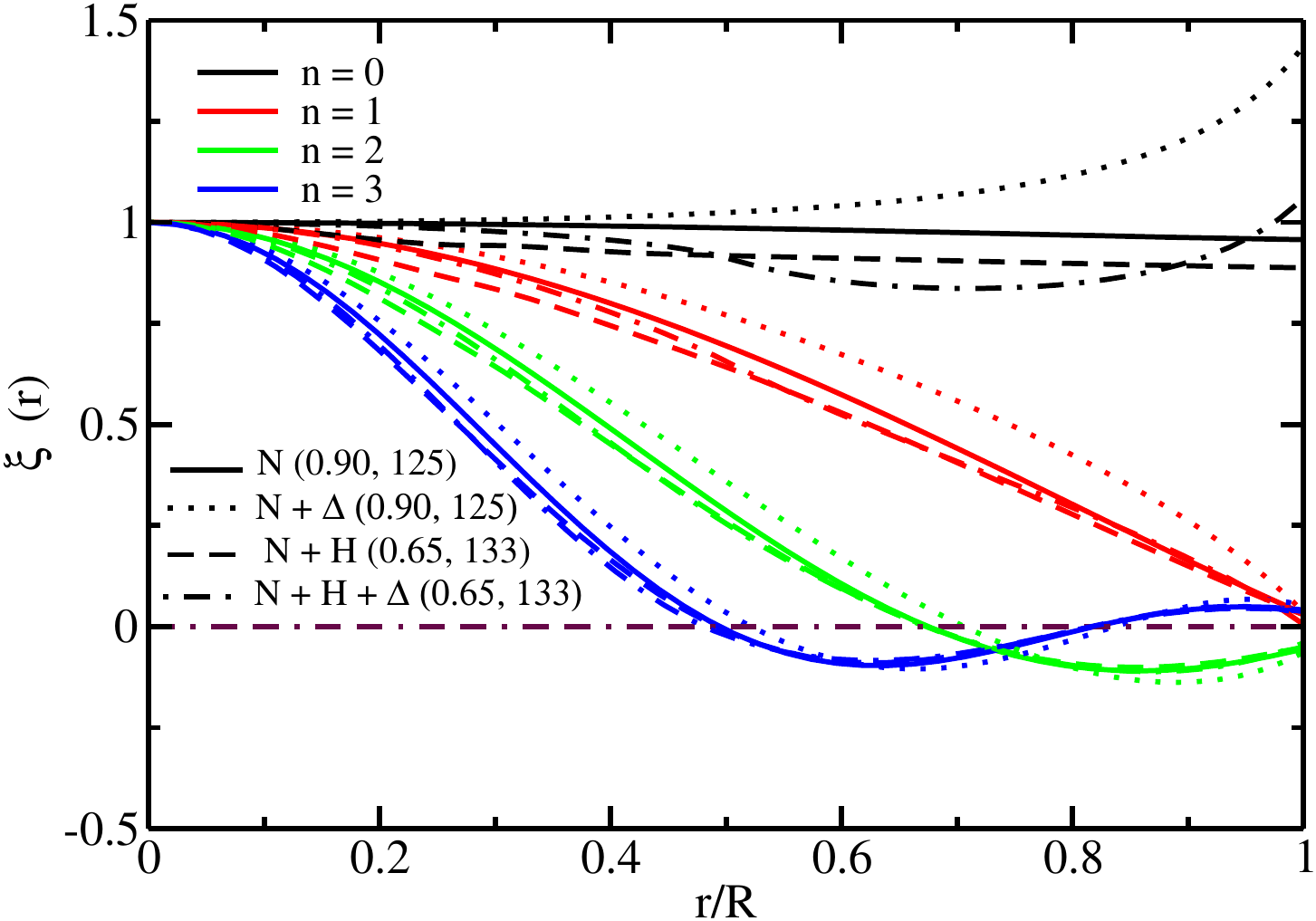}
			 	\end{minipage}
			 			\caption{ The radial displacement perturbation ${\xi(r)}$ = ${\Delta r/r}$ (right panels) and the radial pressure perturbation ${\eta(r)}$ = ${\Delta r/r}$ (left panels) as a function of dimensionless radius distance ${r/R}$ for lower ${f}$-mode (n = 0), lower order ${p}$-modes (n = 1, 2, 3). }
		\label{figmodes}	 	
     \end{figure*}
     
As shown in Figure \ref{figmodes} right panel, $\xi_n(r)$ amplitudes are larger near the star’s center and diminish towards the surface. Lower modes have smoother profiles, while higher modes display oscillations that intensify with increasing mode number. The fundamental $\xi_0$ profile for N+$\Delta$ and N+H+$\Delta$ shows higher amplitudes near the surface compared to N and N+H. The $\eta$ profiles are more compact and the fundamental $\eta_0$ profile generally trends toward negative surface amplitudes for hybrid EoSs. SSHS radial profiles closely resemble those of stable NSs but exhibit reduced amplitudes and frequencies due to their higher mass and unique composition, particularly influenced by strange quark matter. These profiles indicate structural similarities between SSHS and NSs, with subtle differences reflecting the distinct nature of SSHS.

\begin{figure*}[t]
		\begin{minipage}[t]{0.49\textwidth}		 		
  \includegraphics[width=\textwidth]{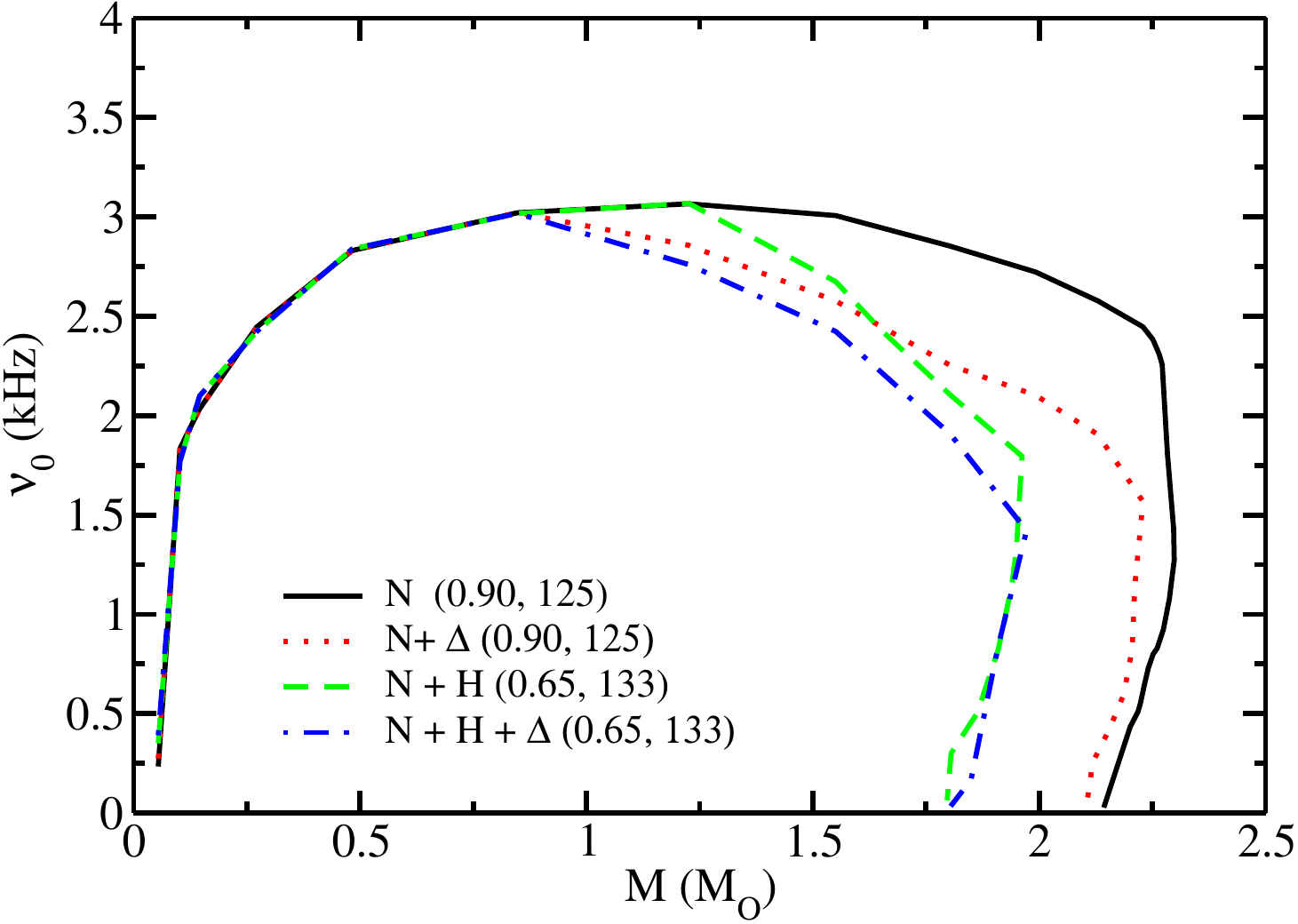}
			 	\end{minipage}
		 		\begin{minipage}[t]{0.49\textwidth}
			 		\includegraphics[width=\textwidth]{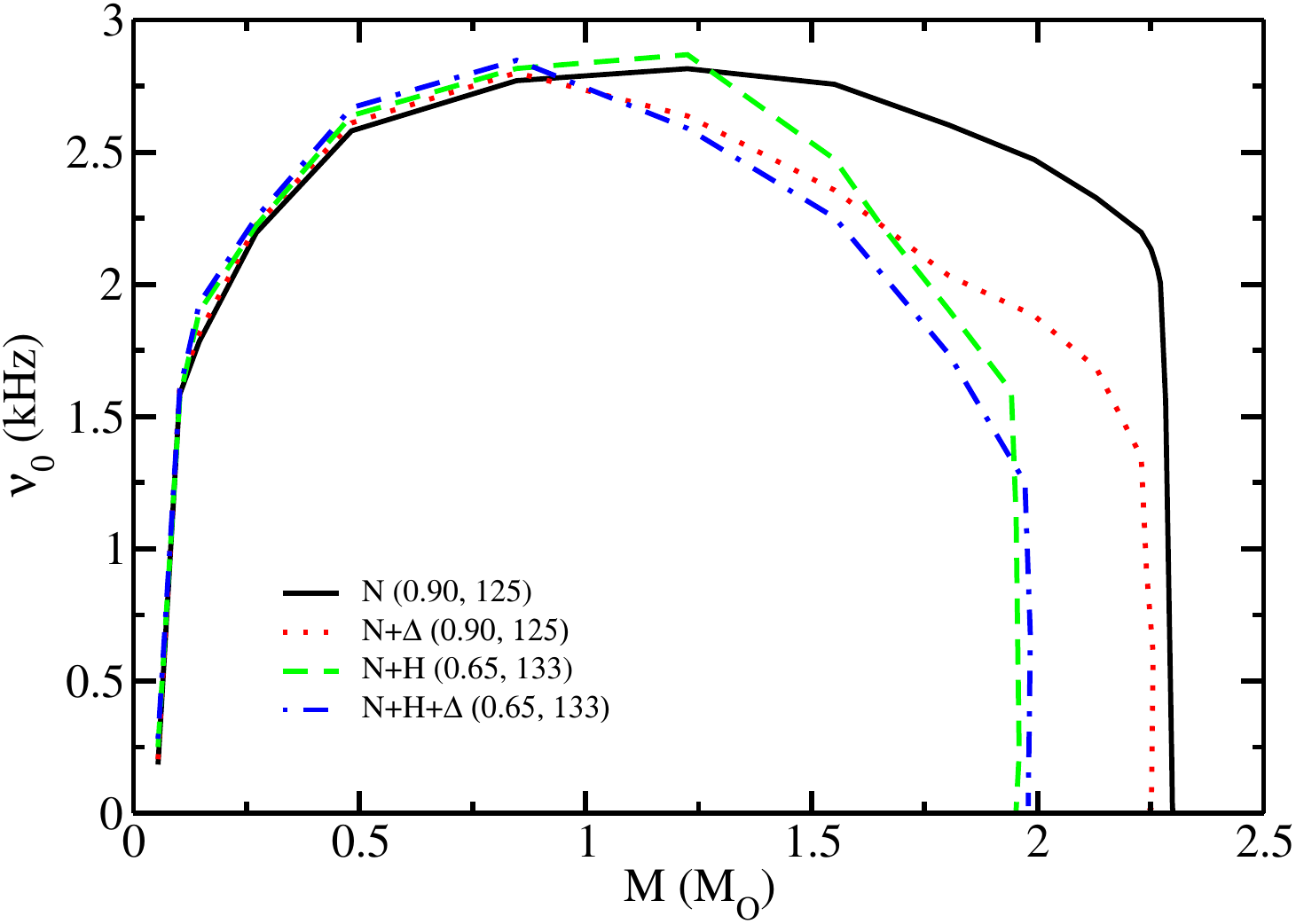}
			 	\end{minipage}
			 			\caption{Fundamental frequency as a function of the mass sequence for slow conversion (left plot) and rapid conversion (right plot) for different compositions of EoS with phase transition. }
		\label{figfreq}	 	
     \end{figure*}

Figure \ref{figfreq} displays the fundamental $f$-mode frequency for slow conversion (left plot) and rapid conversion (right plot) as a function of NS mass for different matter compositions. The fundamental frequency initially increases with stellar mass but starts decreasing near 1\,$M_{\odot}$. As the central density of a star rises, the $f$-mode frequency approaches zero, coinciding with the star nearing its stability limit. However, for slow phase transitions, the maximum mass is reached before the $f$-mode frequency vanishes, suggesting that configurations with central densities beyond the maximum mass can remain stable against small radial perturbations. For the rapid conversion, we see that the frequency drops to zero as soon as the maximum mass is reached. More details on the EoS, the oscillation modes at higher masses, and higher-order frequency modes can be found in Ref. \cite{Rather:2024hmo}.

\section{Conclusion}
\label{conc}
Radial oscillations of hybrid stars were analyzed by solving perturbation equations to determine ${f}$- and ${p}$-mode frequencies and eigenfunctions at 1.4\,$M_{\odot}$. For N+$\Delta$ and N+H+$\Delta$, fundamental profiles ($\xi_0$) showed increased amplitude near the surface, contrasting with N and N+H profiles. Compact $\eta$ profiles were observed at 1.4 $M_{\odot}$, while amplitudes decreased at higher masses. Slow phase transitions allowed stability in configurations exceeding maximum mass. SSHS profiles resembled traditionally stable NSs but showed reduced radial mode amplitudes and frequencies. Mode separations decreased with $\Delta$ baryons and fluctuated for hybrid EoS with hyperons at higher masses. Future studies will incorporate effects such as temperature, rotation, and magnetic fields for a more realistic analysis.
\acknowledgments

I. A. R. acknowledges support from the Alexander von Humboldt Foundation. K. D. M. was supported by the Conselho Nacional de Desenvolvimento Científico e Tecnológico (CNPq/Brazil) under grant 150751/2022-2. B. C. B. was supported by STFC through her PhD studentship under grant ST/W50791X/1. I. L. acknowledges the Funda\c c\~ao para a Ci\^encia e Tecnologia (FCT), Portugal, for the financial support to the Center for Astrophysics and Gravitation (CENTRA/IST/ULisboa) through the grant Project~No.~UIDB/00099/2020 and grant No. PTDC/FIS-AST/28920/2017.



\providecommand{\href}[2]{#2}\begingroup\raggedright\endgroup
\end{document}